# On the question of the asymmetry between matter and antimatter

Jose N. Pecina-Cruz
Department of Physics and Geology
The University of Texas-Pan American
Edinburg, Texas 78501-2999
jpecina@intelligent-e-systems.com

**Abstract**

In this article the lack of equilibrium between matter and antimatter is elucidated. Heisenberg uncertainty principle is a crucial ingredient to understand this disproportion.

**Introduction**

It has been an interesting subject of conversation among physicists, the question of why matter is in a major proportion in the universe than antimatter. A rational answer, based on the principles of quantum mechanics is proposed in this article. Section 1 discusses the relation between matter and antimatter. The conclusion that comes up from section 1 is that matter, must predominate over antimatter if the universe is ruled by quantum mechanics and the theory of relativity.

**1. Matter and Antimatter**

In the one particle scheme Feynman and Stückelberg interpreted antiparticles as particles moving backwards in time [1]. This argument is endorsed by S. Weinberg, who realizes that the antiparticles existence is a consequence of the violation of the principle of causality in quantum mechanics [2]. The temporal order of the events is distorted when a particle wanders in the neighborhood of the light cone. How is the antimatter generated from matter? According to Heisenberg uncertainty principle, a particle wandering in the neighborhood of the light-cone (see Fig. 1) suddenly tunnels from the timelike region to the spacelike region; in this region the relation of cause and effect collapses. Since if an event, at $x_2$ that is observed by an observer A, to occur later than one event at $x_1$ (that is $x_2^0 > x_1^0$.) An observer B moving with a constant velocity **v** respect to observer A, will see the events separated by a time interval given by

$$x_2'^0 - x_1'^0 = L_\alpha^0(v)(x_2^\alpha - x_1^\alpha), \tag{1}$$

where $L^{\beta}_{\alpha}(v)$ is a Lorentz boost. From equation (1), it is found that if the order of the events is exchanged for the observer B, that is, $x'^{0}_{2} < x'^{0}_{1}$ (the event at $x_1$ is observed later than the event at $x_2$.), then a particle that is emitted at $x_1$ and absorbed at $x_2$ as observed by A, it is observed by B as if it were absorbed at $x_2$ before the particle were emitted at $x_1$. The temporal order of the particle is inverted. This event is completely feasible in the neighborhood of the light-cone, since the uncertainty principle allows a particle to tunnel from time-like cone section to space-like cone region. That is, the uncertainty principle consents that the time-like interval to reach values greater than zero as is shown below. Classically the time-like region of the space-time cone is defined by [2],

$$(x_1 - x_2)^2 - (x^0_1 - x^0_2)^2 < 0. \tag{2}$$

But according to quantum mechanics (Heisenberg uncertainty principle),

$$(x_1 - x_2)^2 - (x^0_1 - x^0_2)^2 \leq \left(\frac{\hbar}{mc}\right)^2,$$

where $\left(\dfrac{\hbar}{mc}\right)^2 > 0,$ (3)

and $\dfrac{\hbar}{mc}$ is the Compton wave-length of the particle. The left hand side of the equation (3) can be positive or space-like for distances less or equal than the square of the Compton wavelength of the particle. Therefore, causality is violated. The only way of interpreting this phenomenon is assuming that the particle absorbed at $x_2$, before it is emitted at $x_1$, as it is observed by B, it is actually a particle with negative energy and certain charge (quantum numbers) moving backward in time; that is $t_2 < t_1$ [2]. This event is equivalent to see an antiparticle moving forward in time with positive energy, and opposite charge (quantum numbers) that it is emitted at $x_1$ and it is absorbed at $x_2$. With this reinterpretation the principle of causality is recovered. According to the discussion above, a particle has to violate the postulate, of the special theory of relativity, of the constancy of the speed of light to become an antiparticle. But, this violation of this postulate must occur to maintain the consistency of the principles of quantum mechanics. The infringement -of the speed of light as an upper limit of propagation of a signal- happens in a very short time period according to the uncertainty principle. Therefore, the recombination of matter and antimatter is immediate. The Compton wavelength is a very small distance (see ref. 2) for elementary particles and extremely short for higher masses. That is the probability for crossing the barrier, between time-like and space-like region, decreases when the mass of the particle increases. Therefore, the formation of antimatter (matter moving backward in time) is not a frequent event. In fact antimatter occurs by tunneling between the future time-like cone and the past time-like region.

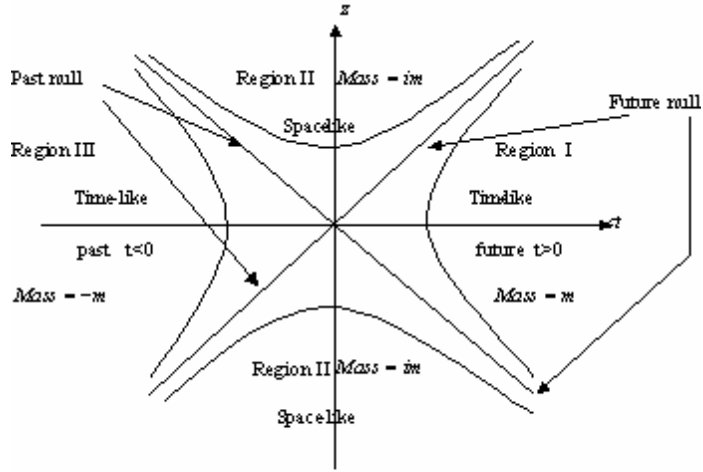

Figure 1

**Conclusion**

The proposed mechanism for the creation of antimatter in Ref. 2 leads to the conclusion that the creation of antimatter is an energetically unfavorable event. From here, in the present energetic state of the universe matter is more abundant than antimatter. Perhaps in a certain moment, after the big bang, the universe was in a state of equilibrium between matter and antimatter. At the present time, the universe is cold enough to allow that matter to be in a major proportion than antimatter.

**Acknowledgments**

I would like to thank to Roger Maxim Pecina for his worthy commentaries to the manuscript. I am in debt to Cris Villarreal for encouraging me to write this article.